\documentclass[twocolumn,aps,superscriptaddress]{revtex4-1}

\usepackage[pdftex]{graphicx}
\usepackage{amsmath}
\usepackage{hyperref}

\begin{document}

\title{Generation of a wave packet tailored to efficient free space excitation
of a single atom}

\author{Andrea Golla} 
\affiliation{Institute of Optics, Information and Photonics,
  University of Erlangen-Nuremberg, 91058 Erlangen, Germany}
\affiliation{Max Planck Institute for the Science of Light, 91058
  Erlangen, Germany}

\author{Beno\^{i}t Chalopin}
\affiliation{Max Planck Institute for the Science of Light, 91058
  Erlangen, Germany}
\affiliation{Laboratoire Collisions Agr\'{e}gats R\'{e}activit\'{e} -
  UMR5589, Universit\'{e} Paul Sabatier - B\^{a}t. 3R1b4, 31062
  Toulouse Cedex 09, France}

\author{Marianne Bader}
\affiliation{Institute of Optics, Information and Photonics,
  University of Erlangen-Nuremberg, 91058 Erlangen, Germany}
\affiliation{Max Planck Institute for the Science of Light, 91058
  Erlangen, Germany}

\author{Irina Harder}
\affiliation{Max Planck Institute for the Science of Light, 91058
  Erlangen, Germany}

\author{Klaus Mantel}
\affiliation{Max Planck Institute for the Science of Light, 91058
  Erlangen, Germany}

\author{Robert Maiwald}
\affiliation{Institute of Optics, Information and Photonics,
  University of Erlangen-Nuremberg, 91058 Erlangen, Germany}
\affiliation{Max Planck Institute for the Science of Light, 91058
  Erlangen, Germany}

\author{Norbert Lindlein}
\affiliation{Institute of Optics, Information and Photonics,
  University of Erlangen-Nuremberg, 91058 Erlangen, Germany}

\author{Markus Sondermann}
\email{Markus.Sondermann@physik.uni-erlangen.de}
\affiliation{Institute of Optics, Information and Photonics,
  University of Erlangen-Nuremberg, 91058 Erlangen, Germany}
\affiliation{Max Planck Institute for the Science of Light, 91058
  Erlangen, Germany}

\author{Gerd Leuchs}
\affiliation{Institute of Optics, Information and Photonics,
  University of Erlangen-Nuremberg, 91058 Erlangen, Germany}
\affiliation{Max Planck Institute for the Science of Light, 91058
  Erlangen, Germany}

\date{\today}

\begin{abstract}
We demonstrate the generation of an optical dipole wave
suitable for the process of efficiently coupling single quanta of
light and matter in free space. 
We employ a parabolic mirror for the conversion of a
transverse beam mode to a focused dipole wave and show the required
spatial and temporal shaping of the mode incident onto the
mirror.
The results include a proof of principle correction of the parabolic
mirror's aberrations. 
For the application of exciting an atom with a single photon pulse we
demonstrate the creation of a suitable temporal pulse envelope.
We infer coupling strengths of 89\% 
and success probabilities of up to 87\% for the application
of exciting a single atom for the current experimental parameters.
\end{abstract}

\maketitle

\centerline{\small The final publication is available at 
\href{http://www.epj.org}{www.epj.org}.}

\section{Introduction}
\label{introduction}

In the last decades light-matter interaction has been studied in a large
variety of experiments not only in many body systems such as atomic
ensembles, where coherent laser pulses are transferred and stored to a
collective excitation~\cite{lukin2003,hammerer2010}, but also on the
level of single matter particles.
Efficient coupling of single atoms to single photons has been 
demonstrated by locating an atom inside an optical
cavity~\cite{raimond2001, walther2006, kimble1998, rempe1993}. 
Such efficient coupling is mandatory in many proposals for
quantum information protocols which rely on quantum memories and
processors, i.e., on the efficient interaction
of a quantum state of light with the internal quantum states of matter. 
Light is a fast and robust way to transmit
quantum information, but the transfer over long distances requires
quantum repeaters based on such quantum
memories~\cite{cirac1997,briegel1998,duer1999}.
The rate of information transfer
through the quantum channel depends on the process efficiency of both
storage and read-out. 
Such efficient coupling can be achieved using a cavity based scheme and
accepting the corresponding modification of the mode spectrum from
continuous to discrete. 

An alternative -- and quite natural -- approach is the coupling in
\emph{free space}.
In this approach, the \emph{direct} coupling of light to matter has
to be maximized instead of the coupling to a cavity.
Besides their potential use in various applications, efficient free
space coupling schemes offer large design freedom.
They enable the investigation of light--matter interaction
without modifying the density of states of the electromagnetic field
continuum.
Alternatively one can study the effect of, e.g., a modified quantum
vacuum~\cite{gardiner1986}.

Recently, one could witness increased efforts devoted to the
topic of efficient free space coupling.
When discussing the corresponding advances one can clearly distinguish
between two different processes: 
One is the elastic scattering of a laser beam by an interacting
particle for which no substantial excitation of the matter system
occurs, i.e., the probability to find the matter system in its excited
state is low throughout the interaction process.
The efficient elastic scattering manifests itself in the attenuation
of a weak coherent laser beam resonant with a transition of the
interacting particle, as has been shown for single molecules
\cite{wrigge2008}, single quantum dots \cite{vamivakas2007}, single
atoms \cite{tey2008-np} or single ions \cite{slodicka2010}. 
Another hallmark of efficient elastic scattering is the phase shift of
a weak laser beam detuned from resonance, which was demonstrated
recently for the free space scenario
\cite{aljunid2009,pototschnig2011}. 
Corresponding theoretical contributions can be found in
Refs. \cite{kochan1994,vanenk2001,vanenk2004,zumofen2008,tey2009}.

The other process treated in recent literature is the process of
efficient absorption of a photon by a single atom, i.e., the matter
system is to be brought to its excited state with high probability.
Note that this should not be confused with a $\pi$-pulse.
While a $\pi$-pulse -- typically containing many photons -- transfers
the atom efficiently to the excited state the pulse cannot be fully
absorbed by a single atom.
Furthermore, the rotation angle of the Bloch vector (e.g. $\pi$) is given
by time integrating the electric field envelope, while a single photon
Fock state has a zero electric field expectation value.  
Theoretical and conceptual contributions devoted to the absorption of
single photons can be found in
Refs. \cite{quabis2000,lindlein2007,sondermann2007,pinotsi2008,stobinska2009,wang2011}.
A recent experimental demonstration of the absorption of a single photon by
a single atom was given by Piro et al~\cite{piro2011}, whereas an
early demonstration of reduced transmission of a bright laser beam
by a single ion was performed by Wineland et al~\cite{wineland1987}.

It is common to the two processes of elastic scattering and absorption
of a photon that their efficiencies are proportional to the extent to
which the incident light mode resembles an electric dipole wave
\cite{quabis2000,vanenk2004,lindlein2007,pinotsi2008,zumofen2008}. 
More precisely, the spatial properties of the incident light mode should
match the ones of the wave that would emerge from the dipole moment of the
interacting particle.
Here, we show how to experimentally generate such a dipole wave
envisioning close to optimum coupling efficiencies.
As outlined below, we do this by preparing a plane light wave that, after
reflection off a parabolic mirror, is transformed into a
dipole wave \cite{lindlein2007,sondermann2007}.
An essential property of this setup is that the parabolic mirror
covers nearly the complete solid angle surrounding the dipolar absorber.

For the specific aim of perfectly exciting a single two-level atom
with a single photon one may take guidance from the spontaneous emission
process:
The incident photon must resemble the time reversed version of a
spontaneously emitted photon
\cite{quabis2000,lindlein2007,sondermann2007}. 
This includes the dipole properties discussed above.
Furthermore, time reversal arguments demand that the temporal envelope
of the incident photon is exponentially increasing with a time
constant equal to the lifetime of the atom's excited state
\cite{sondermann2007}.
This heuristic approach has been confirmed by detailed calculations
\cite{stobinska2009}.

There exist several
methods for temporal shaping of single photons in general.
One approach is to shape the photon during its creation process, for
instance by temporal modulation of the intensity of the control laser
in schemes based on electromagnetically induced transparency~\cite{eisaman2005} 
or by modulation of the pump beams driving the process of four wave
mixing \cite{chen2010} or parametric
down-conversion (PDC)~\cite{kalachev2010}. 
Also the shaping of single photons by spectrally filtering one of
two PDC photons has been demonstrated~\cite{haase2009}.
In the case of single photon generation with a single emitter located
inside an optical resonator~\cite{kuhn2002, keller2004, mckeever2004,
  bochmann2008} the temporal modulation of the pump laser intensity
changes the emitted single photon waveform. 
Furthermore a heralded single photon can also be modulated directly
using electro-optic modulation~\cite{kolchin2008,olislager2010}. 

However, with the mentioned techniques it is challenging to generate a
single photon Fock state with an exponentially increasing temporal envelop,
especially at wavelengths in the ultraviolet spectral range (which are of
interest in this paper, see below). 
Instead we use coherent laser pulses which are highly
attenuated to less than one photon per pulse. 
This procedure is often used as a substitute for generating single
photon Fock states~\cite{gisin2002}.     
For the sake of simplicity, we will use the term
'single photon' also for strongly attenuated coherent states, 
keeping in mind that all excitation probabilities mentioned
in this paper are strictly valid only for true single photon wave
packets.
Nevertheless, if the amplitude of the coherent state is low enough,
contributions from Fock states with photon numbers larger than one are
negligible.
It is then a good approximation to apply the excitation probabilities
for true single photon states to that instances of the coherent state
wave packets that would be projected onto a single photon state upon
measuring the photon number. 

In what follows, we present and characterize experimental techniques
to produce the optimal optical mode for perfectly coupling light with
an atomic dipole, especially for the case of exciting a single atom
with a single photon.
The paper is organized as follows:
In the next section we motivate a measure for the coupling strength
between a single atom (or any other matter system) and light and relate
this measure to the absorption efficiency.
In Sec. \ref{Generation_optimum_doughnut} the experimental generation
of a radially polarized Laguerre-Gauss mode is shown. 
We present the characterization of this beam and calculate the overlap
of the experimental data with the optimal dipole wave. 
Another property of dipole waves is their uniform phase.
Therefore, it is essential to compensate for phase aberrations of the
device that focuses the incident light towards the matter particle.
This topic is treated in Sec. \ref{Aberration_correction}.
Section \ref{Temporal_pulse_shaping} is devoted to the shaping of the
temporal properties of a weak coherent state with the aim of
maximizing absorption.
This aspect as well as the spatial mode shaping is exemplified for
two specific linear dipole transitions of singly and doubly ionized
Ytterbium, but the corresponding schemes can be adapted to any dipole
transition. 
The properties of our exemplary transitions are listed in
table~\ref{table_transitions}.
Finally, in the last section we estimate the coupling efficiencies and
absorption probabilities that would be achieved when interfacing the
generated light waves with the corresponding single ions.

\begin{table} 
\begin{tabular}{c|c|c|c|c}
	Ion & Transition $\quad$  & Wavelength ($\lambda$)
        $\quad$ & Lifetime ($\tau$) & Label \\ \hline \hline 
	$^{174}$Yb$^{+}$ & $^{2}$S$_{1/2}\rightarrow\ ^{2}$P$_{1/2}$ 
        & 369.5~nm & 8.1~ns & T1 \\  
	$^{174}$Yb$^{2+}$  & $^{1}$S$_{0}\rightarrow\ ^{3}$P$_{1}$ 
        & 251.8~nm & 230~ns & T2 \\ \hline 
\end{tabular}
\caption{\label{table_transitions}
Overview of the dipole transitions treated exemplarily in this paper.
Lifetime data are taken from Refs.~\cite{berends1993,zhang2001}.
The labels T1 and T2 will be used in the text for the sake of
simplicity.}
\end{table}

\section{Definition of the free space coupling strength}
\label{coupling_strength}

The coupling constant of the interaction of light and matter in the
dipole approximation is given by the Rabi frequency
$g\sim\vec{\mu}\cdot\vec{E}/\hbar$ in free space as well as in an optical
resonator, where $\vec{\mu}=\mu\cdot\vec{e}_\mu$ is the atomic dipole
moment and $\vec{E}$ the electric field.  
Depending on whether the treatment is semi-classical or fully
quantized, $\vec{\mu}$ alone or both $\vec{\mu}$ and $\vec{E}$ are
operators, respectively.
In either case a large coupling constant requires a large portion of the
incident electric field to be parallel to the atomic dipole.

In cavity quantum electro-dynamics the coupling strength is
proportional to $g^2$. 
Here, we adopt this measure (cf. Ref. \cite{sondermann2007}) and
define the free space coupling strength $G$ as the electric energy
density in the focus that corresponds to the field component parallel
to the atomic dipole and normalize it to the maximum possible one for
a given input power: 
\begin{equation}
G= \frac{ |\vec{E}\cdot\vec{e}_\mu|^2}{|E_\textrm{max}|^2}
 \quad .
\end{equation} 

In the focus, the electric field component along a specific dipole
moment is proportional to the square root of the solid angle
covered by the mediating focusing optics $\Omega_\mu$ and to the
overlap $\eta$ of the incident field with the dipole field
distribution $\vec{E}_\mu$ \cite{sondermann2008}:
$\vec{E}\cdot\vec{e}_\mu\sim \sqrt{\Omega_\mu}\cdot\eta$.
It should be noted that $\Omega_\mu$ is the solid angle weighted
with the angular dipole radiation pattern $D(\vartheta)$: 
$\Omega_\mu=\int D(\vartheta)\sin\vartheta d\vartheta d\varphi$.
The overlap
\begin{equation}
\label{eq:eta_spherical}
\eta=\frac{
\int \vec{E}^\star\cdot\vec{E}_\mu \sin\vartheta\ d\vartheta d\varphi
}
{
\sqrt{\int |\vec{E}|^2 \sin\vartheta\ d\vartheta d\varphi
\cdot \int |\vec{E}_\mu|^2 \sin\vartheta\ d\vartheta d\varphi
}
}
\end{equation}
is calculated integrating over the solid angle covered by
the focusing optics \cite{sondermann2008}.
The maximum values of these quantities are
$\Omega_\mu^\textrm{max}=8\pi/3$ and $\eta^\textrm{max}=1$. 
Therefore, the free space coupling strength is written as
\begin{equation}\label{eq:G}
G= \frac{3}{8\pi}\cdot \Omega_\mu \cdot \eta^2 \quad .
\end{equation}
It reaches unity only if the full solid angle is covered by the
focusing optics and the mode matching is perfect.
Likewise the maximum coupling strength for half solid angle
illumination is $G=0.5$.
These values correspond to the 'scattering ratios' of four and two,
respectively, as used by Tey et al \cite{tey2009} and Zumofen et al
\cite{zumofen2008}.
 
$G$ can be interpreted as a geometric quantity that is determined by
the focusing setup and the spatial properties of the incident light mode.
For a classical field or a coherent state wave packet $\sqrt{G}$
scales the field amplitude at the focus (or its expectation value,
respectively).
If the incident light field is a Fock state, $\sqrt{G}$ scales the
magnitude of the uncertainty of the incident field at the focus.

While $G=1$ is necessary for reaching the maximum effect
in all sorts of free space interaction (phase shifts, scattered
power, absorption, etc.) it is not sufficient for reaching unit probability
during the absorption of a single photon by a single atom. 
As mentioned above, the temporal profile of the single photon wave
packet plays a decisive role.
The optimum effect, i.e., 100\% excitation of a two-level atom can
only be obtained if the temporal profile matches an increasing
exponential with proper time 
constant \cite{sondermann2007,stobinska2009,wang2011}.
To quantify deviations from this profile we introduce the temporal
overlap $\eta_t$ of the incident field amplitude $E_\textrm{inc}(t)$
with the amplitude
$E_\textrm{ideal}(t)=e^{\frac{\Gamma}{2}t}\cdot\theta(-t)$ of the 
ideal pulse shape:
\begin{equation}
\label{eq:eta_t}
\eta_t=\frac{\int_{-\infty}^\infty E_\textrm{inc}(t) \cdot
   E_\textrm{ideal}(t) dt} 
{\sqrt{\int_{-\infty}^\infty |E_\textrm{inc}(t)|^2 dt / \Gamma}}
\quad .
\end{equation}
$\theta(t)$ is the Heaviside function and $\Gamma$ the
spontaneous emission rate of the two-level system to be excited.
This overlap can be considered as a generalization of Eq. (8) of
Ref. \cite{stobinska2009} which treats the case of
an exponentially increasing pulse amplitude.
It is assumed that the incident signal is shifted in time by an amount
that maximizes $\eta_t$.
Since the respective equation in Ref. \cite{stobinska2009} determines
the probability amplitude to find an atom in the excited state, 
we take $\eta_t^2$ as a multiplicative factor determining the absorption
probability 
\begin{equation}
\label{eq:Pabs}
P_a= G \cdot \eta_t^2 = \frac{3}{8\pi}\cdot \Omega_\mu \cdot \eta^2
\cdot \eta_t^2 \quad .
\end{equation}

The prospective coupling efficiencies $G$ and absorptions
probabilities $P_a$ based on the experimentally achieved overlaps will
be given in Sec. \ref{discussion}.

\section{Generation of the optimum spatial mode}
\label{Generation_optimum_doughnut}

A dipole wave incident from nearly the full solid angle is
generated by reflection of a collimated laser beam from a deep
parabolic mirror. 
The ideal beam shape incident on the parabolic mirror is the time
reversed version of a dipole wave that emerges from the focus of the
mirror and is reflected at the parabolic surface.
After reflection off the parabolic mirror the field vector
$\vec{E}_\mu$ of light emitted by a linear dipole 
located at the focal point and with the quantization axis along the
mirror axis is described by~\cite{lindlein2007, sondermann2008} 
\begin{equation}
\label{eq:dipole}
\vec{E}_\mu\left( \dfrac{r}{f} \right) = 
E_{0}\cdot
\frac{
  \dfrac{r}{f}}{ \left[ \left( \dfrac{r}{2f} \right) ^2 +1 \right] ^2}
\cdot\vec{e}_{\text{r}} 
\end{equation}
where $f$ is the focal length of the mirror, $\vec{e}_{\text{r}}$ the
unit vector in radial direction, $E_0$ is an amplitude factor and a
factor $\exp(i\vec{k}\vec{r})$ has been omitted. 
This field distribution has to be generated by an optical element from
an incoming plane wave.
The amplitude distribution of this ideal mode can in
principle be created using diffractive optical elements. However,
there are several drawbacks to this method~\cite{lindlein2007}.
Instead we generate a radially polarized doughnut beam which can have 
a large overlap with the ideal mode and can be created without using
diffractive optics. 

The field vector $\vec{E}_{\text{rpd}}$
of a radially polarized doughnut mode can be expressed as 
\begin{equation} \label{eq_rad_pol_doughnut}
\vec{E}_{\text{rpd}}\left( \dfrac{r}{f}\right)=E_0\cdot
\dfrac{r}{f}
\cdot e^{-\frac{(r/f)^2}{(w/f)^2}} \cdot \vec{e}_{\text{r}} \quad. 
\end{equation}
The overlap of the radially polarized doughnut mode with the ideal
distribution is optimized by adjusting the beam radius $w$ with the
maximum achievable overlap depending on the magnitude of the weighted
solid angle $\Omega_\mu$ (i.e., the depth of the mirror with respect
to its focal length or the half opening angle) \cite{sondermann2008}.
Our parabolic mirrors have a focal length of $f=$2.1~mm and a front
aperture radius of 10~mm.
This corresponds to a length of $5.67\cdot f$ (aperture radius
$r_\textrm{max}/f=4.76$, half opening angle of 134$^{\circ}$) and a
weighted solid angle for a linear dipole of
$\Omega_\mu$=0.94$\cdot\frac{8\pi}{3}$.
The maximum possible overlap for this geometric parameters is
$\eta$=0.982 for a beam radius $w=2.26\cdot f$ \cite{sondermann2008}.  
This limits the experimentally achievable coupling strength $G$ and
absorption probability $P_a$ to 0.91.
A bore of 0.75~mm radius concentric to the optical axis of the
mirror (cf. Sec. \ref{Aberration_correction}) induces only negligible
changes to $\Omega_\mu$ and $\eta$.

\begin{figure}
\begin{center}
\includegraphics{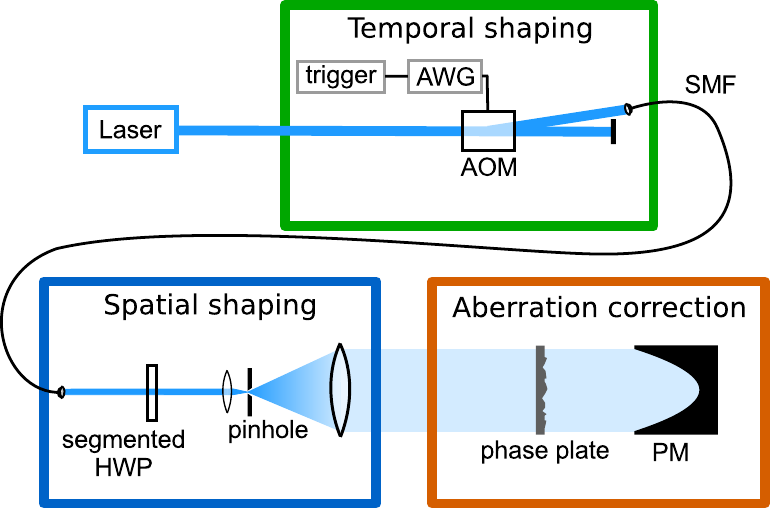}
\caption{\label{fig1}
  Experimental setup.
  The temporal profile is formed  using an arbitrary waveform
  generator (AWG) which determines the temporal envelope of the light
  power scattered into the first diffraction order of an acousto
  optical modulator (AOM).
  The output of the AOM is filtered by a single mode fiber (SMF).
  Linearly polarized light is sent through a segmented half wave plate
  (HWP).
  Subsequently the produced doughnut mode is filtered by a pinhole
  aperture and expanded to appropriate dimensions.
  The aberrations of the parabolic mirror (PM) are compensated using
  a phase plate.
}
\end{center}
\end{figure}

There are different methods for producing a radially polarized doughnut
beam. Examples are the use of sub-wavelength dielectric
gratings~\cite{bomzon2002,ghadyani2011}, nematic liquid
crystals~\cite{stalder1996} and the coherent addition of the two modes
TEM$_{01}$ and TEM$_{10}$ inside laser cavities~\cite{oron2000} or
interferometers~\cite{tidwell1990, tidwell1993}. 
We use a segmented half-wave plate similar to the one
described in Refs.~\cite{dorn2003,quabis2005} to generate the radially
polarized doughnut beam. 
It consists of $n=8$ segments  of a low order
half-wave plate suitable for both wavelengths (370~nm and 252~nm). 
The experimental setup to generate the doughnut mode is shown in
Fig.~\ref{fig1}. 
A continuous wave laser beam at the wavelength of interest first
passes the setup for temporal pulse shaping
(cf. Sec. \ref{Temporal_pulse_shaping}). 
It is then sent through a polarization maintaining single mode fiber
(370~nm beam) or a filtering telescope with a pinhole aperture (252~nm
beam), respectively.
The resulting linearly polarized Gaussian beam is centered onto the
segmented half-wave plate where its polarization is rotated in every
segment by an angle of $2\pi/n$, obtaining a full $2\pi$ rotation of
the electrical field vector over the whole beam cross section.
After propagation through the segmented half-wave plate the beam contains
the radially polarized doughnut mode and modes of higher order. 
The latter ones are filtered out using a pinhole aperture in the
focal plane of a Keplerian telescope. 
With this setup the doughnut mode is generated with a power
efficiency of 70\%. 

\begin{figure*}
\begin{center}
\includegraphics{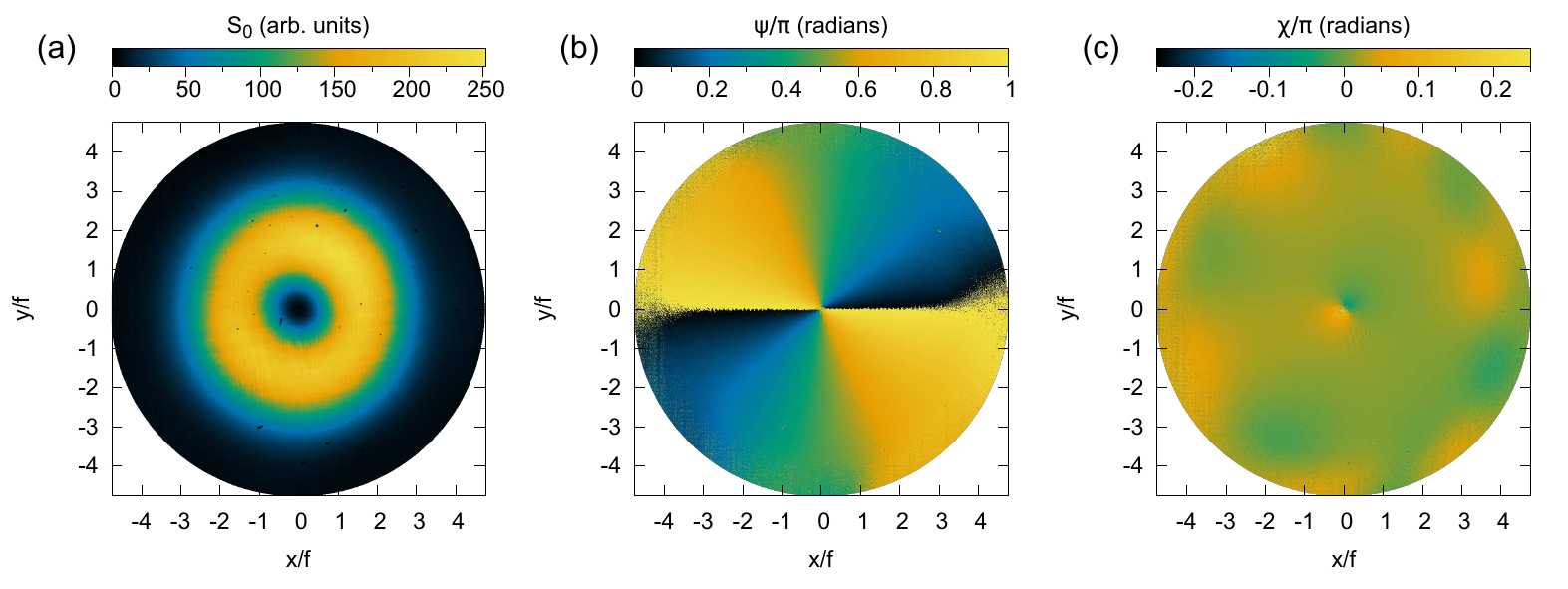}
\caption{ \label{fig2} 
  Characterization of the radially polarized doughnut beam at
  370~nm (transition T1): 
  (a) Stokes parameter $S_0$ (total intensity). 
  (b) Orientation angle $\psi$. 
  (c) Ellipticity angle $\chi$.
  The shown data points cover the aperture of the parabolic mirror.
} 
\end{center}
\end{figure*} 
 
\begin{figure*}
\begin{center}
\includegraphics{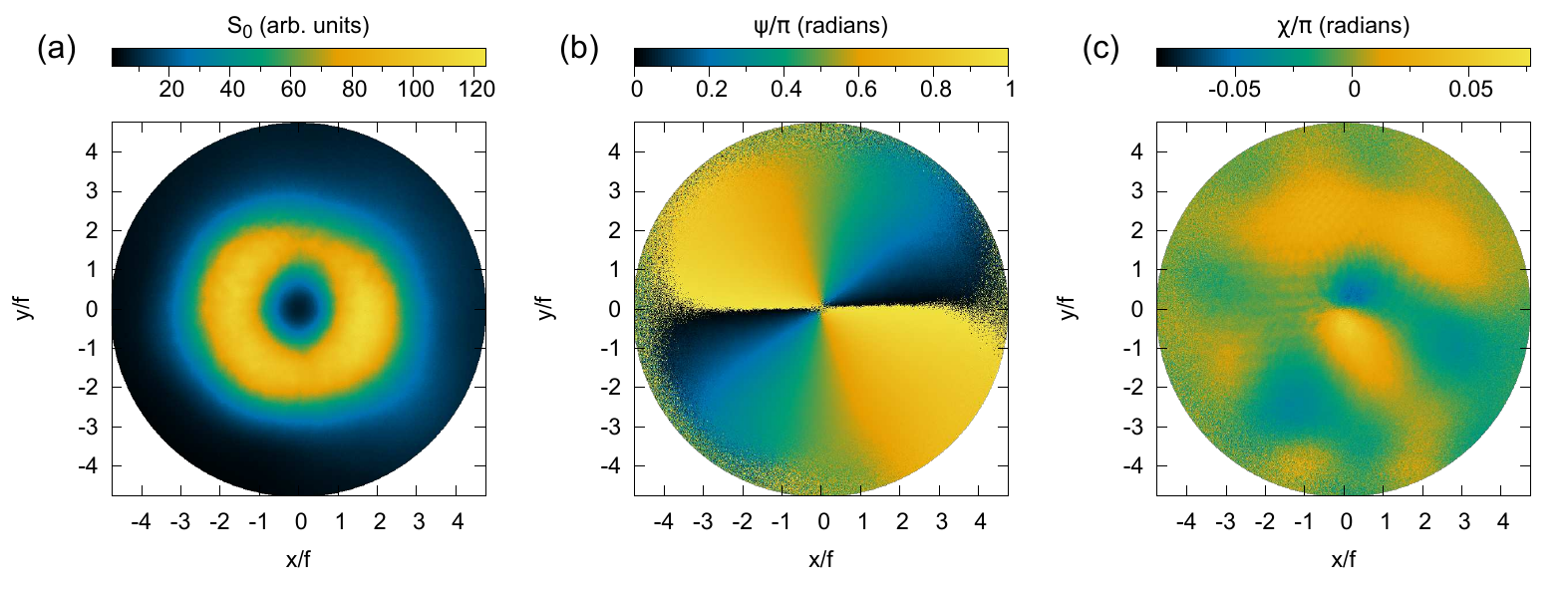}
\caption{ \label{fig3}
  Characterization of the radially polarized doughnut beam at
  252~nm (transition T2):
  (a) Stokes parameter $S_0$ (total intensity). 
  (b) Orientation angle $\psi$. 
  (c) Ellipticity angle $\chi$.
  The shown data points cover the aperture of the parabolic mirror.
} 
\end{center}
\end{figure*}

The polarization properties of the generated doughnut beam are
determined performing spatially resolved  measurements of the Stokes
parameters based on the method described in Ref. \cite{schaefer2007}.
The profile of a beam sent through a rotatable quarter-wave plate and a
non rotatable linear polarizer is measured with a CMOS camera.
Due to the technical limitations of the CMOS camera the Stokes
measurements are not performed with single photon pulses but
with continuous wave laser light. 

Once the Stokes parameters are measured the total intensity $S_0$ and
the characteristic angles of the polarization ellipse are
calculated for every camera pixel.
Since the CMOS camera chip used in the Stokes measurements is much
smaller than the front aperture of the parabolic mirror, the
measurements have been performed before the final magnification step. 
Nevertheless, the data are presented in a coordinate system scaled
such that the beam radius of the measured 
doughnut mode matches the optimum beam radius predicted by theory. 
The results are shown in Figs.~\ref{fig2}
and~\ref{fig3} for 370~nm and 252~nm, respectively. 

The orientation angle $\psi \in \left[0 ,\pi\right]$ describes the
rotation of the main axis of the polarization ellipse with respect to
the x-axis. 
As shown in Figs.~\ref{fig2}b) and
\ref{fig3}b), at both wavelengths $\psi$ covers the full range
of $0$ to $\pi$ in the lower and the upper half plane of the beam
cross section. 
The value of $\psi$ changes continuously with rotation around the beam
center.
The relative phase of the electric field is the same in both half
planes, otherwise a defect line would be observed in $S_0$ at the
$y=0$ line where $\psi$ jumps from zero to $\pi$.
Hence, the orientation of the polarization vector is radial on the
majority of the beam cross section.
Deviations from this behavior occur mainly for radii $\gtrsim 3.5\cdot
f$, i.e., in regions of low intensity.
We attribute these deviations to measurement noise originating from the
discretization of the camera pixels' intensity values.
Furthermore, measurement noise is particularly cumbersome in the
vicinity of the $x$-axis, since there it leads to fluctuations between
values of $\psi=0$ and $\psi=\pi$.
When calculating the overlap with an ideal, noise-free state of
polarization (performed below) these fluctuations lead to pixels on
which the scalar product of the field vectors becomes negative.
For pixels with a large value of $S_0$ this results in a 
measurement-induced reduction of the overlap integral.

The second angle describing the polarization ellipse is the
ellipticity angle $\chi\in[-\pi/4,\pi/4]$ giving the amount of
elliptical polarization. 
Figs. \ref{fig2}c) and \ref{fig3}c) show
the ellipticity angles over the cross section of the generated
doughnut beams.
On the majority of the beam cross section the ellipticity angle is
rather low, i.e., the local state of polarization can be regarded as
linear with a radial orientation.

Next we calculate the overlap of the ideal electric dipole field and
the generated radially polarized doughnut mode in the entrance plane
of the focusing parabolic mirror, assuming perfect reflection at the
mirror surface.
This gives the same value as
calculating the mode overlap on a sphere close to the focal point, as
described in Ref.~\cite{sondermann2008}. 
The integration is performed over the mirror aperture, i.e., only
pixels lying within the boundary $r\le r_\textrm{max}=4.76\cdot f$ are
contributing. 
The spatial field overlap $\eta$ of the measured radially
polarized doughnut beam with the ideal dipole distribution reflected
at the mirror is calculated by taking the normalized scalar product of
the electric fields
\begin{equation} \label{eq_overlap_doughnut_dipole}
\eta=\dfrac{\int E_\text{meas}^\star \cdot 
  \vec{e}_\text{meas}^\star \cdot \vec{E}_\mu r d r
  d\phi }{\sqrt{\int \vert E_{\text{meas}}\vert^{2} r d r d\phi
    \int\vert \vec{E}_\mu\vert^{2} r d r d\phi}} \quad , 
\end{equation}
where $\vec{e}_{\text{meas}}$  is the normalized measured polarization
vector constructed from  the angles $\psi$ and $\chi$. 
$E_{\text{meas}}$  is calculated  by taking the square root of the
measured intensity distribution (the Stokes parameter $S_0$) and 
$\vec{E}_\mu$ is given by Eq. (\ref{eq:dipole}).
Equation (\ref{eq_overlap_doughnut_dipole}) is the equivalent to
Eq. (\ref{eq:eta_spherical}).

For the mode driving transition T1 we calculate the field overlap of the
generated radially polarized doughnut mode with the ideal dipole wave
to be 0.963.
If we assume perfect orientation angles for the measured beam, i.e.,
we disregard the discretization effects discussed above, we find
$\eta_\textrm{370}=0.979$.
The overlap of the generated doughnut mode for transition T2 with the
ideal dipole mode is calculated to be 0.939. Assuming ideal
orientation angles we arrive at $\eta_\textrm{252}=0.975$.  

Our parabolic mirrors are made from aluminum.
One might wonder whether the radially varying amplitude of the
reflection coefficient, which is due to the  radially varying angle of
incidence, might influence the obtainable overlaps (see also
Sec. \ref{Aberration_correction}). 
Accounting for the radial variation of the reflectivity in
calculating the overlap of the modes given by Eqs. (\ref{eq:dipole}) 
and (\ref{eq_rad_pol_doughnut}) leads to a slight increase by 0.03\%
in comparison to the maximum obtainable mode overlap for constant
reflectivity.
  
Therefore, we conclude that the obtainable overlaps are not
influenced by the material properties of aluminum in a significant way.

\section{Aberration correction}
\label{Aberration_correction}

An essential prerequisite for generating dipole modes is an
aberration free phase front:
The phase of the focused wave should be uniform on a
spherical surface enclosing the focal point.
Such an ideal spherical wave can only be generated if the parabolic
mirror is free of surface deviations. Since any realistic device has
aberrations (see, e.g.,
Refs.~\cite{leuchs2008,drechsler2001,stadler2008}), these need to be
characterized and corrected for. 
Ideally the wavefront incident onto the parabolic mirror can be shaped
such as to create an aberration free inward moving spherical wave
after reflection off the mirror. 

One method of measuring the imperfections of a parabolic mirror is to
scan its surface using a
profilometer~\cite{stadler2008,drechsler2001}. But the huge relative
depth of our parabola makes it hardly accessible for profilometer
heads. 
We measure the shape deviations by
interferometry with nanometer accuracy~\cite{leuchs2008}. 
Let us briefly recall the core features of the interferometer
setup, details can be found in Ref.~\cite{leuchs2008}.
We use an interferometer of the Fizeau type at the standard
helium-neon laser wavelength $\lambda_{\text{HeNe}}=632.8$~nm.  
A reference wave is produced at a wedged fused silica plate with
surface flatness $\le\lambda/$100. 
A small steel (or glass) sphere located concentric to the focus of the
parabolic mirror is acting as a null element.
Any light ray traveling parallel to the optical axis of the parabola
is reflected towards the sphere, reflected back onto its
path of incidence and interferometrically overlapped with the
reference beam. 
Owing to this double pass geometry, we measure twice the phase
aberration introduced by the mirror. 
Possible aberrations of the sphere are eliminated by averaging over
different spheres and different rotational orientations of each sphere
relative to the parabolic mirror.
The deviations from a perfect parabola obtained after removal of
misalignment aberrations are finally represented by Zernike
polynomials.

Our parabolic mirrors are made of diamond turned aluminum and
are produced at the Fraunhofer Institute for Applied Optics and
Precision Engineering at Jena (see
Sec.~\ref{Generation_optimum_doughnut} for the geometric parameters). 
A central on axis bore of 1.5~mm diameter is used for locating the
sphere or any other object such as an ion~\cite{maiwald2009} in the
focal point. 

The interferometric measurements are performed with a radially
polarized doughnut beam centered on the optical axis of the parabolic
mirror. 
In this particular case light is polarized parallel to the 
plane of incidence at every spot on the surface of the parabolic reflector.

\begin{figure*}
\centering
\includegraphics{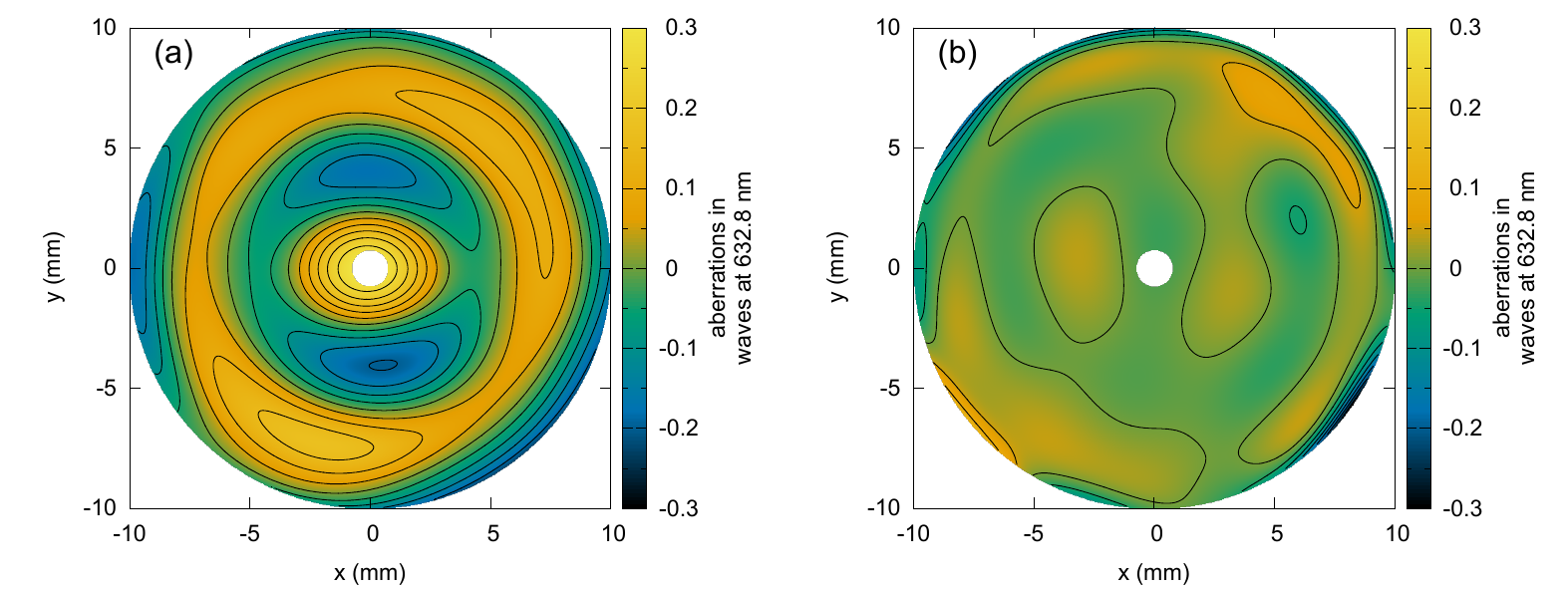}
\caption{
\label{fig4}
(a) Measured phase aberrations in single pass reflection. The
resulting aberrations are shown averaged over two different 
spheres and six rotational settings per sphere. 
The PV value is 0.53 waves. 
(b) Measured aberrations with a compensating phase plate (single
experimental run). 
The PV value is 0.45 waves.
In both panels the data result from a Zernike fit of degree 10.
Adjacent contour lines are separated by 0.05 waves at
$\lambda_\textrm{HeNe}$.  
}
\end{figure*}

Figure~\ref{fig4}a) shows the result of such a measurement
for one of our parabolic mirrors.
The peak-to-valley (PV) value of the aberrations is 0.53 waves at
$\lambda_{\textrm{HeNe}}$. 
The root mean square (RMS) value is 0.09 waves at
$\lambda_{\textrm{HeNe}}$.
The data shown correspond to maximum deviations of about 130~nm from
the ideal parabolic shape~\cite{leuchs2008}.
This is of the same order of magnitude as reported for the parabolic
mirror used in Ref.~\cite{drechsler2001}.

Based on these interferometry results we have produced a
phase plate which nominally imprints the phase conjugate of the
measured aberrations onto an otherwise flat phase front.
The fused silica phase plate is produced in house by laser lithography
and (selective) reactive ion etching. 
In order to check the quality of aberration compensation the phase
plate is inserted into the interferometric setup between the parabolic
mirror and the flat reference plate.
While constantly repeating the interferometric measurement, position and
orientation of the phase plate are optimized.
As a first optimization criterion we take the reduction of the
PV value of the aberrations while for the fine adjustment the RMS
value is a better measure, since this is the quantity determining the
Strehl ratio~\cite{born-wolf1991}.
The final outcome of such a measurement series is depicted in
Fig.~\ref{fig4}b).
The PV value of the aberrations is reduced to 0.45 waves at
$\lambda_{\text{HeNe}}$. 
The RMS value shrinks to 0.033 waves at $\lambda_{\text{HeNe}}$.
It should be noted that the seemingly small reduction of the PV value
is due to effects occurring at the boundaries of the phase plate and
the parabolic mirror.  
Neglecting the outer 0.5 mm of the aperture the PV value shrinks to 
0.16 waves at $\lambda_{\text{HeNe}}$ and the RMS value is reduced to
0.023 waves. 

To judge the quality of these improvements we have performed
simulations of the focal intensities based on a generalization of the
method by Richards and Wolf~\cite{richards1959}. We assume a wave
front according to the aberrations shown in Fig.~\ref{fig4}
to be incident on a perfectly reflecting parabolic mirror.
The incident light wave is a radially polarized doughnut mode (see
Eq.~(\ref{eq_rad_pol_doughnut})) with the optimum beam radius.
The resulting focal intensities are normalized to the focal intensity 
obtained for an aberration free wave front. 
This yields the Strehl ratios.
At the testing wavelength of our interferometer (633~nm), the Strehl
ratio is improved from 68\% to 99\% by using the correcting phase
plate. 
In other words, the combined system 'parabolic-mirror-plus-phase-plate'
is practically capable of diffraction limited focusing.

At the wavelengths for experiments with Ytterbium ions the Strehl
ratios of the uncorrected mirror are 32\% for 370~nm and 9\% for
252~nm. 
Close to diffraction limited performance seems feasible in
these cases if the phase plate is designed for the respective
wavelength considering the corresponding refractive index
of fused silica.

Besides that, a phase plate that perfectly compensates the aberrations  
at the testing wavelength $\lambda_{\text{HeNe}}$ is also rather
suitable at the wavelengths of the transitions T1 and T2. 
Assuming a phase plate working perfectly at the testing wavelength, we
simulated the performance of the compound system
'phase-plate-plus-paraboloid' at the other two wavelengths.
Since the relative refractive index changes of fused silica at the
wavelengths of interest are on
the order of 1\% -- 3\%, our simulations reveal Strehl ratios ranging
from 97\% to more than 99\%, depending on the wavelength and the
specific parabolic mirror.

So far we have assumed a phase shift upon
reflection that does not depend on the distance to the mirror's
optical axis. 
But due to the complex refractive index of aluminum this assumption
is generally not justified. 
The radially varying phase shift upon reflection imposes a defocus. 
Nevertheless we argue that the influence of the refractive index of 
aluminum can be neglected. 
We calculate the phase shift upon reflection by taking the respective
refractive indices from Ref. \cite{smith1998} and determining the
argument of the complex reflection coefficient for electric fields
parallel to the plane of incidence.
For this we use the relations between angle of incidence and
radial distance to the optical axis given in
Ref.~\cite{lindlein2007}. 
The obtained phase shifts are then used in simulations with an
otherwise aberration free paraboloid.
For the wavelengths considered here the simulations show an axial
shift of the intensity maximum by less than a tenth of a wavelength. 
The maximum intensity is reduced by amounts on the order of 0.2\% in
the case of transition T2 and even less for the other wavelengths.
The intensity in the nominal focus is reduced by
0.5\% (1.5\%, 2.8\%) at the testing (transition T1, transition T2)
wavelength.

The phase shifts induced by the above effects can in principle be
measured by our interferometric setup.
However, a substantial part of the treatment of the interferometric
data is the removal of misalignment aberrations of the null sphere
from the exact focal position~\cite{leuchs2008}.
Since the monotonous radial variation of the phase shift upon
reflection is equivalent to a slight displacement of the null sphere,
this effect is removed in the measurement procedure. 

We finish this section by noting that the orders of magnitude of the
experimental results shown are generally reproduced for other
parabolic mirrors, with the shape of the aberrations of the
uncorrected mirrors differing from device to device.

\section{Generation of the optimum temporal mode}
\label{Temporal_pulse_shaping}

To design the temporal mode of a single photon wave packet that
would be absorbed efficiently by a two level system, we exploit the
time reversal symmetry of the compound system atom and free space
field modes~\cite{sondermann2007}. An excited two level atom
spontaneously decays into the ground state under emission of a photon
with a probability amplitude that is exponentially decreasing in time
\cite{weisskopf1930}. 
If a photon wave packet with such a 
decreasing envelope is sent onto a two level system the maximum
possible excitation probability is  calculated to be
54\%~\cite{stobinska2009}.

An excitation probability of 100\% is achieved
by generating a photon wave packet with the time reversed shape of the
spontaneously emitted photon: an exponentially increasing field
amplitude with a time constant which is twice the lifetime of the
excited state. 
Furthermore the ideal pulse length is infinite and has a sharp edge at
the end. 
The excitation probability depends not only on a good overlap of the
time constant and the sharp edge at the end of the pulse, but also on
the total pulse length. In Ref.~\cite{stobinska2009} it is shown
theoretically that for pulse lengths of five lifetimes 99\% excitation
probability can be achieved. 
These numbers are valid for single photon Fock states. 
However, as outlined already in Sec.~\ref{introduction}, coherent state
pulses with average photon numbers $\ll$1 are a good approximation to
single photon wave packets.

\begin{figure}
\begin{center}
\includegraphics{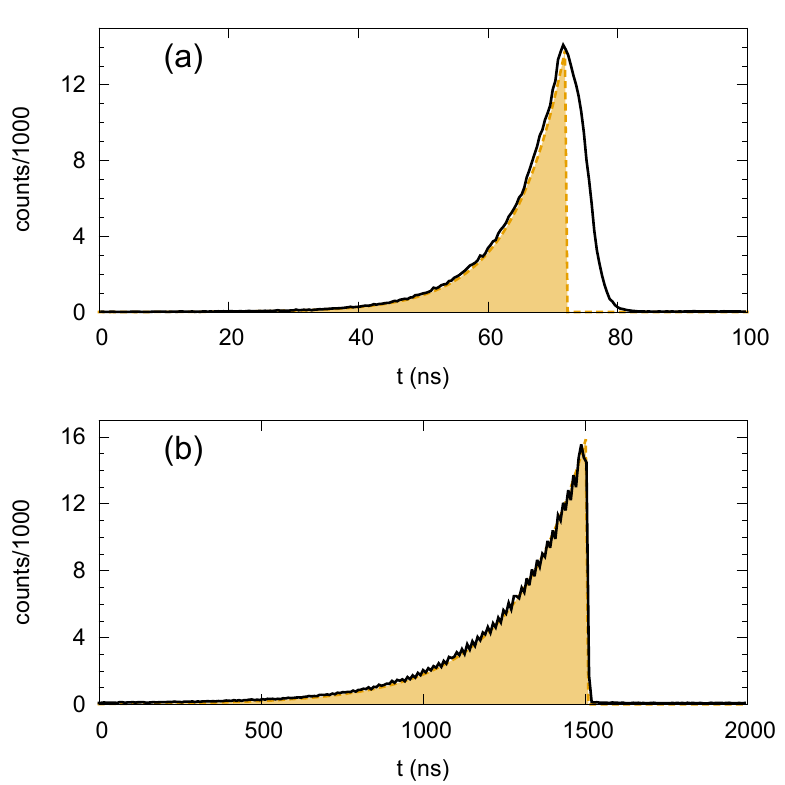}
\caption{ \label{fig5}
  Generated exponentially increasing pulses with a mean photon
  number per pulse of 0.09~photons. The dashed line indicates the
  perfect pulse shapes. The time constant (life time) is 8~ns in panel
  (a) and 230~ns in panel (b).
  The total number of counts are $430\cdot10^3$ and
  $487\cdot10^3$, respectively. } 
\end{center}
\end{figure}

In the following, we discuss the generation of these coherent state wave
packets.
The experimental apparatus for the generation of arbitrarily shaped
pulses is shown in Fig. \ref{fig1}. 
A continuous wave laser beam is cut into pulses using an acousto-optic
modulator (AOM). 
The intensity diffracted into the first order of the AOM depends on
the amplitude $U_0$ of the applied radio frequency (RF) signal as 
$I\left(t\right)\sim\sin^{2}(U_0(t))$. 
An exponentially increasing wave packet is obtained by modulating the
amplitude of the applied RF signal $U_{\text{RF}}(t)=U_0(t)\cdot
\sin(\omega_\text{RF} t)$ as   
\begin{equation}
U_0(t)=\arcsin\left(\exp\left[\dfrac{t}{2\tau}\right]\right)
\end{equation}
where $\omega_\textrm{RF}=2\pi\cdot400$~MHz is the RF driving
frequency of the AOM. 
The modulated RF signal is generated with a digital arbitrary waveform
generator and amplified before being applied to the AOM. 
The first diffraction order of the AOM is coupled to a polarization
maintaining single mode fiber and sent to the setup for the spatial beam
shaping. 
The attenuation to a mean photon number per pulse of much less
than one is done using a set of half-wave plates and linear
polarizers. 
The temporal profile of these highly
attenuated laser pulses is measured using a photomultiplier tube with
a quantum efficiency of 15\% and a frequency counter in start-stop
configuration. 
Since we do not measure a photon in every pulse, the start trigger is
given by the detection of a photon and the stop trigger by the
following clock signal activating the output of the arbitrary waveform
generator. 
We account for this trigger configuration by reversing the
temporal coordinate of the experimental data.
In Fig.~\ref{fig5} the measurements of the pulses designed to match
the lifetimes of transition T1 and T2 are shown. 
The mean photon number per pulse is 0.09.
For simplicity reasons the 370~nm laser was used for the creation
of the pulses for both time constants. 

The temporal overlap $\eta_{\text{t}}$ of the field amplitudes is
calculated according to Eq. (\ref{eq:eta_t}).
The incident field amplitude $E_\textrm{inc}(t)$
is obtained by taking the square root of the photon count statistics
and the integration is performed over the time intervals covered by
the experimental data shown in Fig. \ref{fig5}.
The results of this procedure are $\eta_t=$0.96 for the pulse
with 8~ns time constant and $\eta_t=$0.99 for the 230~ns pulse. 
The lower overlap value in the case of the 8~ns pulse is attributed to
the finite build-up/decay time (5~ns) of the optical grating inside
the AOM crystal.
This results in non-negligible deviations from a sharp pulse end when
compared to the time constant of the rising exponential.

\section{Discussion}
\label{discussion}

\begin{table} 
\begin{tabular}{l|c|c}               
	Transition  & T1 & T2 \\ \hline \hline
	Spatial overlap $\eta$ & 0.979 & 0.975  \\ \hline
        Solid angle $\Omega_\mu/\frac{8\pi}{3}$ & 0.94 & 0.94 \\ \hline  
	Strehl ratio  & 0.99 & 0.99 \\ \hline 
	Temporal overlap $\eta_t$& 0.96 & 0.99 \\ \hline \hline
        Coupling strength $G$ & 0.892 & 0.885 \\ \hline
        Absorption probability $P_a$ & 0.812 & 0.867  \\ \hline
\end{tabular}
\caption{ \label{estimation_overlaps}
Overview of the achieved figures of merit.
 The Strehl ratio is taken from the proof of principle experiment at
 $\lambda_\textrm{HeNe}$.
}
\end{table}

With the acquired experimental data at hand we are now able to give an
estimate of the coupling efficiency and absorption probability which
would be achieved in experiments employing the generated field modes.
Table~\ref{estimation_overlaps} summarizes the field overlaps of the
generated light modes in time and in space domain, as well as the
Strehl ratio.
According to Eqs. (\ref{eq:G}) and (\ref{eq:Pabs}) we calculate $G$
and $P_a$ from these values and let the Strehl ratio enter as a
multiplicative factor accounting for losses caused by the 
mirror imperfections remaining after compensation.
We thus arrive at prospective coupling strengths of $G=0.892$
($G=0.885$) for transition T1 (T2).
These values are only about two percent below the upper limit set by the
experimental geometry when using doughnut modes.
The corresponding absorption probabilities are $P_a=0.812$
($P_a=0.867$).

These absorption probabilities are valid under the assumption of a
closed two-level system (e.g., transition T2 of $^{174}$Yb$^{2+}$).
As an example for a more complex level structure we take the
transition T1 of $^{174}$Yb$^{+}$:
Both the excited and the ground state are split into two
Zeeman sub-levels with the magnetic quantum numbers 
$m=\pm 1/2$ (we neglect the dipole allowed
$^{2}$P$_{1/2}\rightarrow\ ^{2}$D$_{3/2}$ transition due to the
small branching ratio of 0.5\% \cite{migdalek1980}). 
Two linear and two circular dipole transitions are possible. 
The probability for spontaneous decay via a linear dipole
transition is $1/3$ according to the corresponding Clebsch-Gordan
coefficient and $2/3$ via a circular dipole transition.
Exciting the linear dipole transition leads to a
maximum possible excitation probability of $1/3$, assuming the
excitation light mode is ideal.  
In this sense, one has to understand the value $P_a=0.812$ given
for a mode designed for this transition as a relative value, i.e., the
de facto absorption probability would be $P_a/3$.
However, the coupling strength $G$ itself is not affected by such
considerations.
For example, an elastic scattering experiment on any of these
transitions should be feasible with the coupling strength derived for
an ideal two-level system because no dipole moments are induced for
transitions other than the driven one (cf. Ref. \cite{jin2011} for a
calculation applying the level structure discussed above). 

We conclude this paper by emphasizing that the method presented here
to generate a dipole mode can be applied to a dipole transition at any
optical wavelength.  
An application to circular dipole transitions is
straightforward requiring corresponding amplitude and polarization
distributions.
However, these polarization patterns become quite complex -- especially
when focusing from a large solid angle \cite{sondermann2008}. 

In summary, we have demonstrated the generation of modes of the
electromagnetic field allowing for efficient coupling of single
photons and single two-level systems in free space using a
deep parabolic mirror. 
The free space coupling method described here should compete
favorably with other coupling schemes.

\acknowledgments
The authors acknowledge fruitful discussions with 
A. Villar, S. Heugel, M. Fischer and J. Schwider.
M.S. acknowledges financial support from the Deutsche
Forschungsgemeinschaft (DFG).


%

\end{document}